# OPTIMIZATION OF A CLASSICAL STAMPING PROGRESSION BY MODAL CORRECTION OF ANISOTROPY EARS

Y. Ledoux*, H. Favrelière, S. Samper

SYMME, Polytech'Savoie, BP 806, F74016, Annecy Cedex
\* corresponding author: Yann.Ledoux@univ-savoie.fr

## Abstract

This work is a development from the INETFORSMEP European project. It was proposed to realize a global optimization of a deep drawing industrial progression (made of several stages) for a cup manufacture. The objectives of the process were the thickness decrease and the geometrical parameters (especially the height). This paper completes this previous work in the aim of mastering the contour defect.

From the optimal configuration, we are looking for minimizing the needed material and the number of forming operations. Our action is focused on the appearance of undesirable undulations (ears) located in the rim of the cups during forming, due to non-uniform crystallographic texture. These undulations can cause a significant amount of scrap, productivity loss and undesired cost during manufacture. In this paper, this phenomenon causes the use of five forming operations for the cup manufacture. The focus is to cut down from five to two forming stages by defining an optimal blank (size and shape). The advantage is to reduce the cost of the tool manufacturing and to minimize the needed material (by suppressing the part flange). The envisaged approach consists in defining a particular description of the ears' part by modal decomposition and then simulating several blank shapes and sizes generated by discrete cosine transformation (DCT). The use of a numerical simulation for the forming operation and the design of experiment technique allows to find mathematical links between the ears' formation and the DCT coefficients. An optimization is then possible by using mathematical links. This original approach leads to reduce the ears' amplitude by ten, with only fifteen numerical experiments. Moreover, we have downsized the number of forming stages from five to two with minimal material use.

# Introduction

This work succeeds to a previous publication where it was realized an optimization of a stamping part formed through few stages [1]. It was applied a global optimization of the whole process. The objectives were:
- to find the minimal blank dimension which leads to assure the flange cutting and the final height of the part.
- to determine the best tool set configuration (tool radii, blank holder force, …) which leads to a minimal thickness decrease along the part.

The stamping progression was composed of 4 stages: cutting the initial blank, the forming operation, the flanging and the cutting operation. These later operations were necessary due to the anisotropy of the material. Because of this non-uniform crystallographic texture, the undesired undulations (ears) appearances were located on the rim of the cups during forming.

This study starts from this optimal tool set configuration where process conditions are optimized. The focus is to suppress these two last operations by modifying the initial size and shape of the blank.

Classically, to compensate these ears formation, manufacturers have to realize different trials on the forming process by modifying the initial blank shape. This kind of approach needs time, process mobilization for the trials realization and different cutting blank tools.

This paper proposes an alternative solution by piloting the blank geometry by a DCT decomposition (Discrete Cosine Transformation). The stamping part defect is then characterized by a modal decomposition. The different trials are realized on a finite element simulation (Abaqus). The experimental design technique is used to find the influence of the blank modification on the final part shape (ear amplitude and final height). These trials allow to calculate different polynomial relations and it is possible to find the optimal blank shape and size which leads to obtain a good part (good dimension without ears).

## 1. Studied part and previous optimal results

The studied part is a cylindrical cup made in only one stamping operation. This particular part has to be welded to another part (not shown here). This welding process forces a small shape defect on the bottom zone (to ensure the quality of the welding operation). The drawing and the associated geometrical tolerancing is indicated in the figure 1.

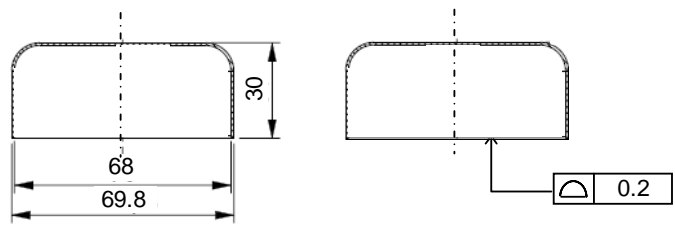

Figure 1 : initial target geometry

The classical industrial part progression is shown in the figure 2. The part comes from a circular blank cut in 0.8 mm thick USB sheet metal (DC05). The different steps are forming, flanging and cutting. Values reported on this figure 2 came from the previous optimization.

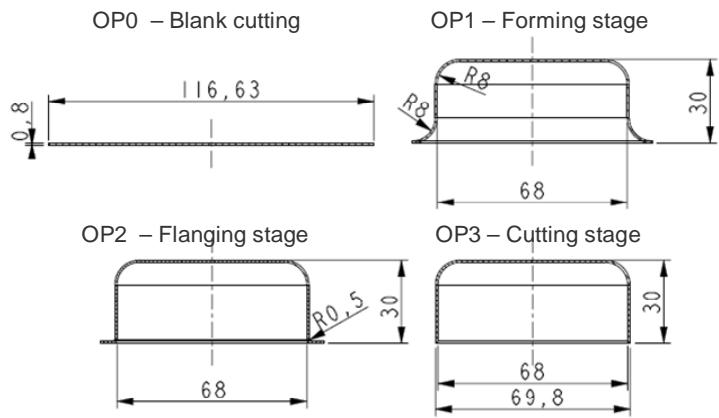
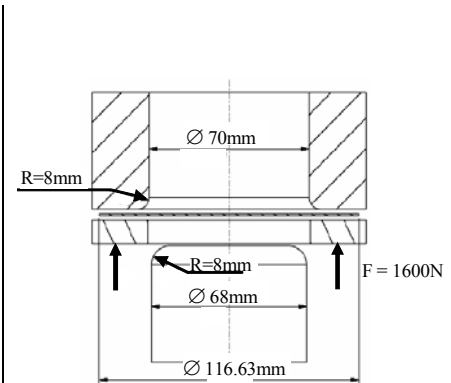

Figure 2: optimal progression of the part.

Figure 3: detail of the OP1 optimal tool set

The focus is then to reduce the progression to only one forming operation (OP0 and OP1 of the figure 2). For that, it is necessary to precisely define the initial shape of the blank. There are two different objectives to respect:
- The final part height is equal to 30mm.
- The ear amplitude does not exceed 0.2mm.

**Assumption:**
In the previous study [1], it is shown that the blank modification have no influence on the minimal thickness value. So we assumed that it is possible to find an optimal blank (size and shape) without influence on thickness value.

### 1.1. Numerical simulation

Abaqus software is used for the numerical simulation in Explicit version [2]. Shell elements (S4R) with four nodes discretize the blank. There are seven integration points in the thickness. The stamping tool is modeled by rigid surfaces. The main parameters used for simulation are listed in table 1.

| Blank discretization | |
|---|---:|
| element type | Shell (4 nodes) |
| integration points | 7 |
| number of elements | 1428 |
| **Tool discretization** | |
| tool type | Rigid surface |
| **Process parameters** | |
| Punch speed | 3m/s |
| friction coefficient | 0.1 |

Table 1. Numerical parameters of the finite element simulation.

The used material is a USB steel sheet (DC05) of 0.8mm thickness. The hardening is determined on tensile tests using extensometer and image analysis system for high strain levels [3]. It is introduced point by point in the FE code. The used behavior law takes into account the anisotropy according to Hill's 48 criterion [4]. The material characteristics are shown in table 2.

| USB (DC05) | |
|---|---:|
| Young modulus | 206.62GPa |
| Yield Strengh | 168.12MPa |
| Poisson modulus | 0.298 |
| Density | 7800kg/m3 |
| Anisotropy coefficients (Lankfort) | $r0° = 2.09$ |
| | $r45° = 1.56$ |
| | $r90° = 2.72$ |

Table 2. Main USB material properties.

## 1.2. Realization of the initial simulation

In the aim of identifying the part defect, the initial simulation of the stamping operation is computed. In this case, the initial blank size is determined by a calculation of the needed area.
The possibility to obtain the wished geometry of the part and the minimal flange is directly linked to the initial blank diameter. So we have to estimate its initial diameter D0 by taking into account these following hypotheses:
- H1: the volume of the blank is constant during the stamping operation.
- H2: the thickness is totally constant (the local decrease is compensated by local increase).

A geometrical calculation gives an initial diameter D0 = 116.63mm.

From this nominal blank value and with the optimal tool configuration, we can measure the initial ear amplitude of 1.72mm (see figure 4).

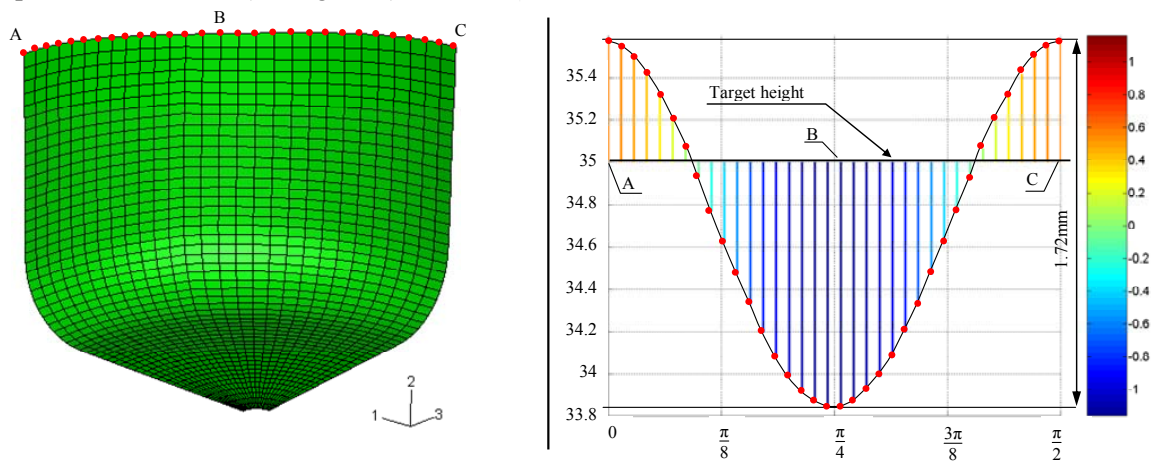

Figure 4 : Initial simulation and corresponding deviations around the target height

## 1.3. Defect type and corresponding setting parameters

### 1.3.1. Classical approaches

In the figure 4, we can see the ear deviation on the part border projected along the axis 2. The graph on the right hand side quantifies the ear amplitude. Each segment represents local deviation of each node around the height target (35mm). There are different ways to drive this kind of defects.
The classical approach consists in associating a theoretical circle on the circumference point. Then, it is possible to define the extreme points that correspond to the shape defect. The data parameters are the position of the circle (point + vector) and a scalar value (distance between the two extreme points). With this approach, we do not know the form defect and it is impossible to identify the corresponding modification to do in the initial blank shape.
Another approach consists in measuring the position of each node and to compare it with the wished position. With these parameters, there are as many distance values as nodes.

### 1.3.2. Discrete geometry parameterization

We propose particular setting parameters by defining a discrete geometry parameterization of the defect. It is then possible to fit the measured geometry with an open basis of derived geometries. Thus size, position and form defects can be decomposed in the following basis (several discrete parameterizations are possible). The first one is Fourier type transformation [7-8] which is well known by the metrologists for the circularity and straightness defects (2D defects for 1 geometry parameter). In the case of optic lenses (ISO 10110-5), Zernike polynoms define a set of possible defects for a disk geometry (3D defects for 2 parameter geometry). The Discrete Cosinus Transformation [9] makes possible to define a set of defects for a rectangular geometry (3D defects for 2 parameter geometry) and is used for image processing.

### 1.3.3. Modal description of the defect

Those solutions are interesting because they allow the description of all the defects according to the parameterization. We propose a model [10-11] which can be used on any geometry. This model is based on modal analysis of the target part geometry.

**Assumptions:**
- Due to anisotropy, the part and the part defects have got two symmetries are symmetrical.
- **Form defects** are taken into account.

**The eigen decomposition modes**

As it is expected to characterize the part profile, 2D beam elements are chosen for this eigen decomposition mode. The clearance between the target and the simulated profile is calculated in a deviation vector *(V)*. The modal analysis is used for computing modal basis $Q_i$.
Then, this basis is used for the decomposition of *V* and for calculating $\lambda_i$ coordinates. These coordinates represent the deviations projected in the modal basis.

**General purpose**

In linear dynamics, the discretized equations of the movement for a conservative, discrete or continuous system can be written in the general form:

$$M\ddot{q} + Kq = 0 \quad (1)$$

In (1), *M* is the generalized mass matrix, *K* is the generalized stiffness matrix and *q* a dynamic displacements vector.
*n* is the freedom degrees number of the system. The solutions are the modes $q_i$ of a structure which can be decomposed as two functions (space*time):

$$q_i = Q_i \cdot \cos(\omega_i t) \quad (2)$$

where $Q_i$ is the amplitude vector, and $\omega_i$ the corresponding pulsations in rad.s$^{-1}$.
Taking into account their definition, the modes are solutions of the equation:

$$(K - \omega_i^2 M)Q_i = 0 \quad (3)$$

(3) admits *n* eigen solutions, which are the structure modes. The modal pulsations $\omega_i$ of the various modes are the roots of the characteristic polynomial:

$$det(K - \omega_i^2 M) = 0 \quad (4)$$

The vectors of modal deformations $Q_i$ are the eigenvectors associated with the pulsations $\omega_i$, which form a basis *M* and *K* orthogonal in the vector space of the structure movements.

The norm is then $||A||\infty = Max(|A_j|)$ \quad (5)

The orthogonality of the eigen modes expresses that the inertias and stiffness developed in a mode do not work in the movements of the other modes. There is mechanical independence of the modes.
This method gives all the possible defects of the part (mode shapes) through modal shape decomposition. In fact, the discretization limits the number of possible defects.

**Application of the method for the studied part**

The stamped part geometry results from numerical simulation.
It is discretized with 35 elements with equal length. The *V* clearance vector is calculated from the deviation of the target geometry to the simulated stamped one. The *eigen modes* of the geometry are identified by modal analysis (see relation 3).
A projection in this base of the *V* vector is done that allows the calculation of the $\lambda_i$ coordinates. Its decomposition is carried out using the vectorial projection:

$$\lambda_i = <V.Qi> = \frac{1}{||Qi||^2} Q_i^{\ t} V \quad (6)$$

Where $\lambda_i$ is the coordinate associated to the mode $Q_i$, so called the modal coordinates. The figure 5 represents the five first mode shapes and can be compared to the target geometry.

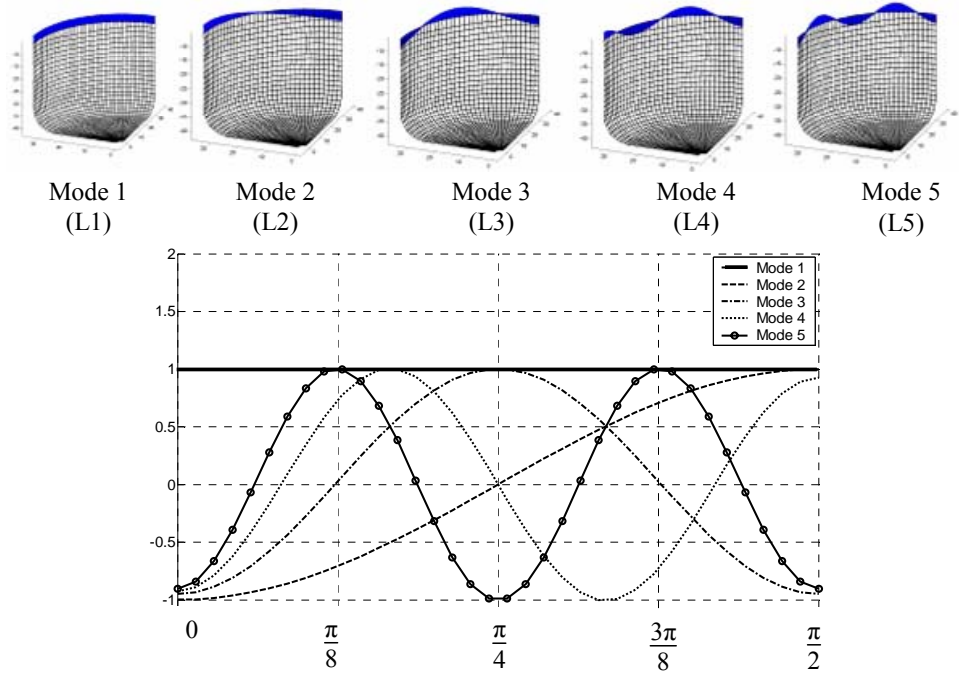

| Mode 1 (L1) | Mode 2 (L2) | Mode 3 (L3) | Mode 4 (L4) | Mode 5 (L5) |

Figure 5 : size and shape modal defects

In this particular case, it is possible to show that the modal decomposition leads to define a discrete cosine shape with different amplitudes and wavelength for the part defect reconstruction.

In this special model, represented in the figure 6, the modal model corresponds to a beam with only one degree of freedom in y axis.

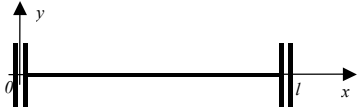

Figure 6: modal model.

It is possible to write then different relations:

$$Y(x) = A_0 Ch(\alpha x) + B_0 Sh(\alpha x) + C_0 cos(\alpha x) + D_0 sin(\alpha x) \qquad (7)$$
$$Y^{(1)}(x) = \alpha (A_0 Sh(\alpha x) + B_0 Ch(\alpha x) - C_0 sin(\alpha x) + D_0 cos(\alpha x)) \qquad (8)$$
$$Y^{(2)}(x) = \alpha^2 (A_0 Ch(\alpha x) + B_0 Sh(\alpha x) - C_0 cos(\alpha x) - D_0 sin(\alpha x)) \qquad (9)$$
$$Y^{(3)}(x) = \alpha^3 (A_0 Sh(\alpha x) + B_0 Ch(\alpha x) + C_0 sin(\alpha x) - D_0 cos(\alpha x)) \qquad (10)$$

*Where $Y^{(i)}$ correspond to the $i^{th}$ function derivate.*

By using the boundary conditions for this model:

$Y^{(1)}(0) = 0,\ Y^{(3)}(0) = 0$
$Y^{(1)}(l) = 0,\ Y^{(3)}(l) = 0$

And then, by using the equations 8 and 10:

$Y^{(1)}(0) = 0 \rightarrow B_0 + D_0 = 0$
$Y^{(3)}(0) = 0 \rightarrow B_0 - D_0 = 0$ $\Big\} B_0 = D_0 = 0$

$Y^{(1)}(l) = 0 \rightarrow A_0 Sh(\alpha l) - C_0 sin(\alpha l) = 0$
$Y^{(3)}(l) = 0 \rightarrow A_0 Sh(\alpha l) + C_0 sin(\alpha l) = 0$

$$\begin{bmatrix} Sh(\alpha l) & -sin(\alpha l) \\ Sh(\alpha l) & sin(\alpha l) \end{bmatrix} \begin{bmatrix} A_0 \\ C_0 \end{bmatrix} = \begin{bmatrix} 0 \\ 0 \end{bmatrix}$$

And finally,

$$\rightarrow \alpha = \frac{n\pi}{l} \rightarrow A_0 = 0$$

And then, Y(x) can be write as follow:

$$Y(x) = C_0 cos\left(\frac{n\pi}{l} x\right) \qquad (11)$$

Figure 7 shows the modal decomposition of the nominal geometry. It can be obvious that the first modes are the most contributing to the modal description of the part's defects.

In this paper, the five first modes are used to describe the part's defects (size and shape). The projection residue value is less than one percent.

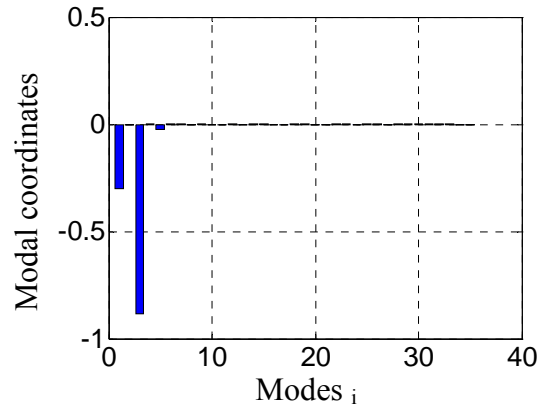

Figure 7 : Initial part defect modal coordinates.

### 1.4. Blank shape control by DCT

For driving the initial size and shape of the blank, a discrete cosine transformation (DCT) is used. We decide to use three different signals. The first (named D) defines the value of the initial blank diameter, the second (named A1), the ovalization (two lobes) and the third (named A3), the four lobe shape. The figure 8 shows the different possible shapes.

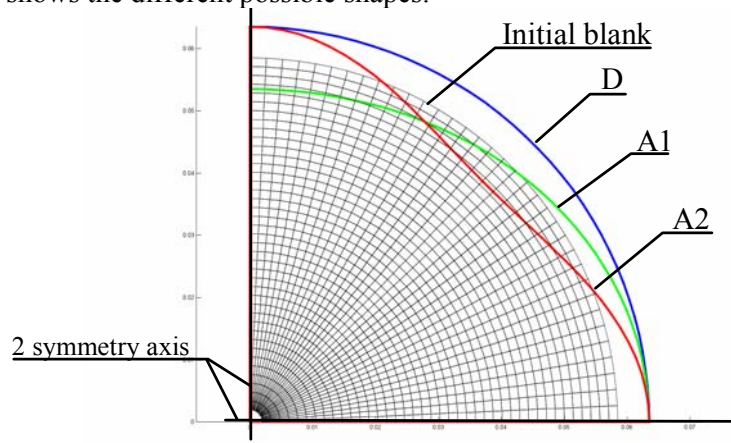

Figure 8 : different blank shapes obtained with D, A1 and A2 parameters.

A combination with these different DCT coefficients allows to obtain on the initial blank more complex shape.

## 2. Optimization strategy

The optimization of the process parameters by the DOE technique needs to follow these four different steps:
1. Selection of the range variation of each process parameters,
2. Choice of the mathematical model and realization of the experiments,
3. Calculation of the polynomial relations and analysis of their influence on the final part defect
4. Optimization by using the obtained mathematical model and verification

### 2.1. Selection of the range variation of each DCT parameters

According to the previous estimation of the initial blank diameter, the initial simulation and its trends, it is possible to define the range variation of each DCT parameters. These choices are very important and could compromise the well running of the optimization.
Two extreme cases can be found. Either the search space is too small then the optimum can be outside,

or the search space is too big and the accuracy can be not sufficient.

We suppose that modifying the lobe shape of ±1.5mm, is sufficient to compensate the anisotropy ears (the initial amplitude corresponds to 1.72mm). Moreover, it seems that the initial blank size is very close to the optimal one because the obtained height with the initial simulation is equal to 34.69mm. The variation range of the diameter is then ±1.5mm.

By taking into account these constraints, we choose their variation ranges.

| Process parameters | Initial value | Low value | High value |
|---|---|---|---|
| D : nominal diameter | 117mm | 115.5mm | 118.5mm |
| A1: 1$^s$t DCT coefficient | 0 | -1.5 | 1.5 |
| A2: 2$^{nd}$ DCT coefficient | 0 | -1.5 | 1.5 |

Table 3: variation range of each DCT coefficient

## 2.2. Model choice and experimental design

We enhanced the method accuracy by selecting a second degree polynomial model which assumes a quadratic variation of the geometrical output from every process input:

$$Y= a_0+a_1X_1+ a_2X_2+\ldots+ a_{12}X_1X_2+ a_{ij}X_iX_j +\ldots+ a_{11}X_1^2 + a_{nn}X_n^2$$

In this model, the terms $X_iX_j$ are interactions between factors, which means that the effect of one of them ($X_i$) depends on the value of the other one ($X_j$).

A Box-Wilson central composite design (called central composite design or CCD design) was used ([5] and [6]). This is a factorial fractional design (experiments 1 to 8 in table 4), with a central point (experiment 9, table 4). It was enhanced with a group of star points to enable estimation of quadratic curvatures. The distance from the center of the design space to a factorial point was ±1 unit (in normalized variation) for each factor, the distance from the center of the design space to a star point was ±α, with |α| > 1. The precise α value depends on some properties desired for the design and on the number of factors involved. In our case, the optimal value for α was ±1.287, [5].

It is possible to define the corresponding experimental design for the quadratic polynomial calculation. Each line of the table 4 corresponds to a numerical simulation carried out with the corresponding blank with particular D, A1 and A2 values. The result was a file of points from the bottom contour. This file was post-processed in order to measure the modal coefficients (L1, L2, L3, L4 and L5) of each simulated part. All these results are presented in table 4.

| N° exp. | D | A1 | A2 | Profile | L1 | L2 | L3 | L4 | L5 | |
|---|---|---|---|---|---|---|---|---|---|---|
| | | | | | 0 | 0 | 0 | 0 | 0 | <- target values |
| 1 | 115.5 | -1.5 | -1.5 | Profile 1 | -1.35E+00 | 1.68E+00 | 8.72E-01 | -3.32E-01 | -9.07E-02 | |
| 2 | 115.5 | -1.5 | 1.5 | Profile 2 | -1.24E+00 | 1.65E+00 | -1.48E+00 | 2.56E-01 | -6.70E-02 | |
| 3 | 115.5 | 1.5 | -1.5 | Profile 3 | -1.22E+00 | -1.03E+00 | 1.12E+00 | -1.29E-01 | -9.22E-02 | |
| 4 | 115.5 | 1.5 | 1.5 | Profile 4 | -1.12E+00 | -7.91E-01 | -1.33E+00 | 8.71E-02 | -8.06E-02 | |
| 5 | 118.5 | -1.5 | -1.5 | Profile 5 | 1.28E+00 | 1.76E+00 | 8.43E-01 | -3.86E-01 | -7.35E-02 | |
| 6 | 118.5 | -1.5 | 1.5 | Profile 6 | 1.34E+00 | 1.74E+00 | -1.59E+00 | 2.78E-01 | -6.45E-02 | |
| 7 | 118.5 | 1.5 | -1.5 | Profile 7 | 1.39E+00 | -9.37E-01 | 1.15E+00 | -1.47E-01 | -1.05E-01 | |
| 8 | 118.5 | 1.5 | 1.5 | Profile 8 | 1.52E+00 | -6.94E-01 | -1.42E+00 | 1.16E-01 | -1.02E-01 | |
| 9 | 117 | 0 | 0 | Profile 9 | 1.82E-01 | 4.48E-01 | -3.02E-01 | -5.16E-02 | -1.99E-02 | |
| 10 | 115.07 | 0 | 0 | Profile 10 | -1.45E+00 | 3.65E-01 | -2.67E-01 | -6.36E-02 | -3.62E-02 | |
| 11 | 118.93 | 0 | 0 | Profile 11 | 1.97E+00 | 4.47E-01 | -3.98E-01 | -9.28E-02 | -4.45E-02 | |
| 12 | 117 | -1.93 | 0 | Profile 12 | 4.32E-02 | 2.00E+00 | -5.24E-01 | -7.16E-02 | -4.03E-02 | |
| 13 | 117 | 1.93 | 0 | Profile 13 | 2.49E-01 | -1.22E+00 | -1.82E-01 | -5.05E-02 | -2.41E-02 | |
| 14 | 117 | 0 | -1.93 | Profile 14 | -1.02E-02 | 2.98E-01 | 1.36E+00 | -3.12E-01 | -1.23E-01 | |
| 15 | 117 | 0 | 1.93 | Profile 15 | 1.02E-01 | 4.67E-01 | -1.74E+00 | 2.37E-01 | -1.52E-01 | |

Table 4: Central composite experimental design.

## 2.3. Calculation of polynomial model coefficients and analysis of their influence

### 2.3.1. Model calculation

From the results of the experimental design, coefficients of the polynomial model linking the modal coefficients to the DCT coefficients were calculated by the multi-linear regression method. The following quadratic equations were obtained (equations 12, 13, 14, 15 and 16):

$$L1 = -1.13*10^{-2} + 7.07*D' + 1.69*10^{-3}*A1' + 4.56*10^{-2}*A2'$$
$$-1.64*10^{-2}*D'^2 - 6.50*10^{-4}*D'*A1' - 1.29*10^{-1}*D'*A2'$$
$$-2.44*10^{-1}*A1'^2 + 7.00*10^{-4}*A1'*A2'$$
$$- 6.84*10^{-1}*A2'^2 \quad (12)$$

$$L2 = -2.89*10^{-3} + 7.56*10^{-4}*D' - 7.56*A1' - 8.53*10^{-4}*A2'$$
$$+ 2.73*10^{-5}*D'^2 + 2.17*10^{-1}*D'*A1' - 8.25*10^{-4}*D'*A2'$$
$$- 4.68*10^{-5}*A1'^2 - 4.12*10^{-4}*A1'*A2'$$
$$+ 6.58*10^{-4}*A2'^2 \quad (13)$$

$$L3 = -3.85 - 1.21*10^{-1}*D' + 7.96*10^{-5}*A1' - 6.99*A2'$$
$$+ 1.46*10^{-1}*D'^2 + 1.25*10^{-5}*D'*A1' + 1.44*10^{-1}*D'*A2'$$
$$-1.16*10^{-1}*A1'^2 - 1.12*10^{-4}*A1'*A2'$$
$$+ 3.21*10^{-1}*A2'^2 \quad (14)$$

$$L4 = 4.42*10^{-4} - 8.42*10^{-5}*D' + 1.69*10^{-1}*A1' - 4.38*10^{-4}*A2'$$
$$- 3.85*10^{-4}*D'^2 + 7.77*10^{-2}*D'*A1' + 3.75*10^{-6}*D'*A2'$$
$$+ 1.44*10^{-5}*A1'^2 - 5.78*10^{-1}*A1'*A2'$$
$$+ 3.85*10^{-4}*A2'^2 \quad (15)$$

$$L5 = - 6.88*10^{-2} + 3.27*10^{-2}*D' - 1.23*10^{-4}*A1' - 6.30*10^{-2}*A2'$$
$$- 2.39*10^{-2}*D'^2 - 5.00*10^{-5}*D'*A1' + 4.80*10^{-2}*D'*A2'$$
$$+ 6.17*10^{-2}*A1'^2 + 7.50*10^{-6}*A1'*A2'$$
$$- 4.46*10^{-1}*A2'^2 \quad (16)$$

Where Li' are the standardized variables, respectively corresponding to Li. They range from –1 to +1, while the corresponding variables range from the minimal to the maximal values.

In order to check this mathematical model, a numerical simulation was carried out at the centre of the domain (number 9 in table 4), corresponding to the mean values of the tool parameters. The simulation results were compared to the mean values of quadratic model for each modal coefficient.

### 2.3.2. Influence of the DCT coefficients on the modal parameters

By using the polynomial relations (equations 12, 13, 14, 15 and 16) we find the most influent parameters (D, A1 and A2) on the modal values (L1 to L5).

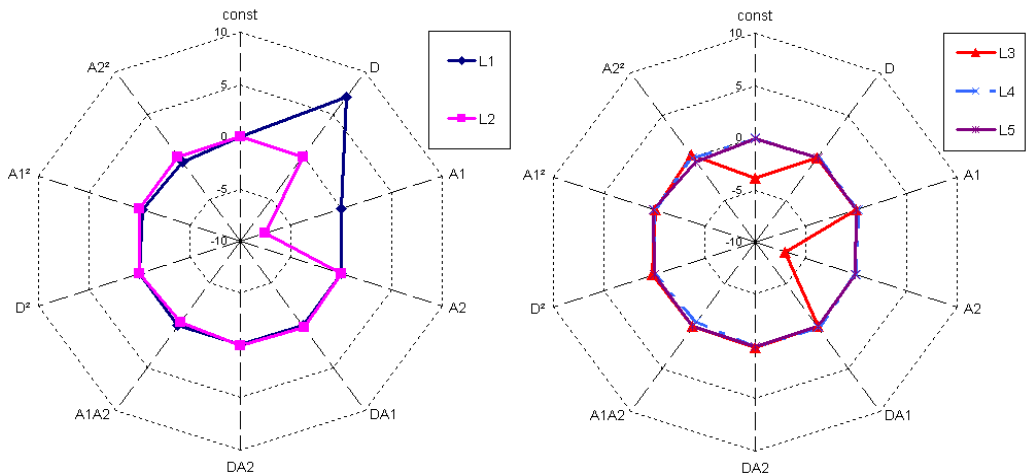

Figure 9: graph of the influence

As it can be seen on the figure 9, the most influent parameter on the L1 defect is D. This parameter directly modifies the initial value of the blank diameter.

The A1 is the most influent parameter on the L2 defect. This parameter drives the ovalization (two lobes along the circumference) of the part and that is L2 which models this defect type.

And then, A2 is the most significant parameter for the L3 defect. A2 allows to give a four lobes form. The defect L4 and L5 defect is quasi equal to zero.

The modal coefficients (Li) are linked to the blank shape (D or Ai). Moreover, we can underline that all the constant terms are null excepted for the L3 coefficient. That means that, with a circular blank, the main defect corresponds to four lobe defects.

### 2.4. Optimal configuration computation

The previous polynomial relations (equations 12, 13, 14, 15 and 16) are used to determine the optimal values of the initial blank which minimizes the shape and size defect of the part.

For that, a particular optimization function is built. And when the function $F$ (eq 17) is minimized, that means that there is no deviation between the target geometry and the stamped part.

$$\text{Minimized } F = \text{Minimized}\left(\sum_{i=1}^{n\ eigen\ mode\ =\ 5} (Li)^2\right) \quad (17)$$

In our case, five modes are used for the defect description, so, the $F$ function (18) is:

$$\text{Minimized } F = \text{Minimized } ((L1)^2+(L2)^2+(L3)^2+(L4)^2+(L5)^2) \quad (18)$$

Optimal values of the initial blank are obtained and a numerical simulation by finite element method is carried out.

| Quadratic optimum | | | | Results | | | | |
|---|---|---|---|---|---|---|---|---|
| D | A1 | A2 | | L1 | L2 | L3 | L4 | L5 |
| 117.05 | -4.51E-04 | -8.07E-01 | Optimum | -0.01 | 8.21E-04 | -2.78E-02 | 7.92E-04 | -8.13E-02 |
| | | | Nominal | -0.30169 | -1.06E-03 | -8.82E-01 | 3.85E-05 | -2.39E-02 |

Table 5: Optimal configuration and modal decomposition of the corresponding part defect.

The corresponding optimum and nominal decompositions are presented in the figure 9.

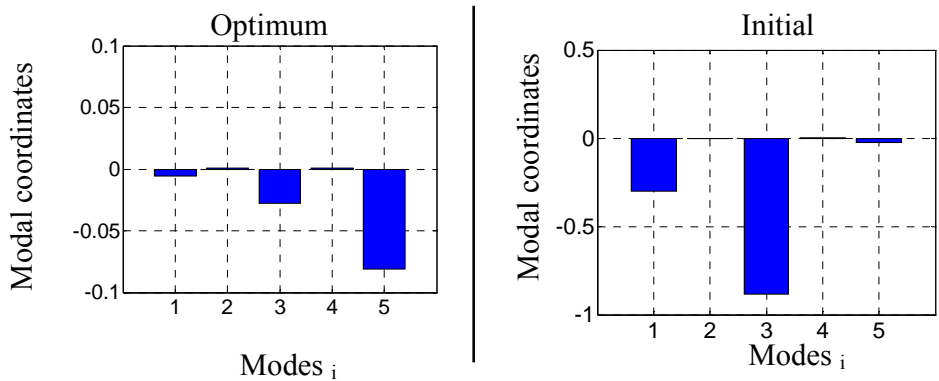

Figure 9: comparison of the optimum and initial defect modal decomposition

The corresponding size and shape defects are represented in the figure 10. With this optimization, the ear amplitude is divided by ten compared to the ears obtained initially.

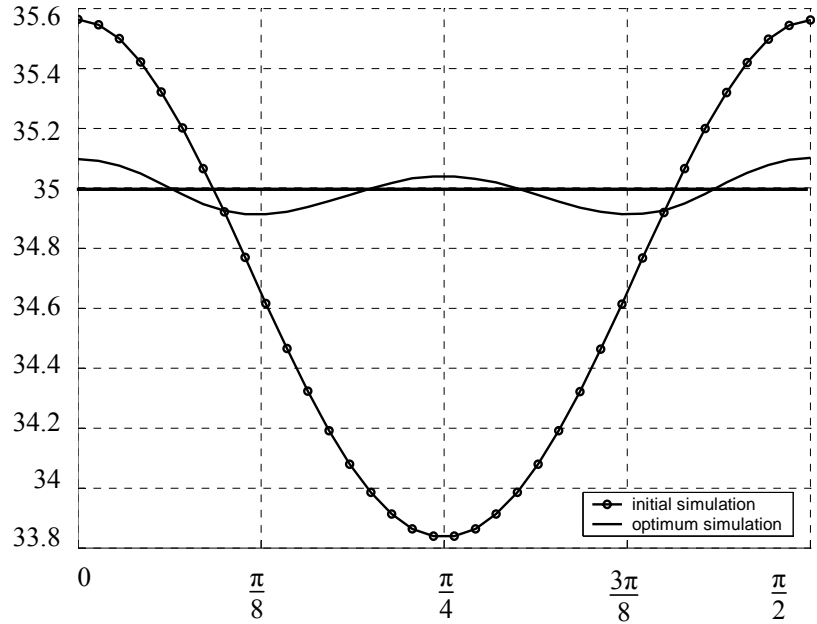

Figure 10: comparison of the nominal and optimum deviations

As we can see, with only fifteen numerical experiments it is possible to improve considerably the blank geometry and then to decrease the number of forming operations (two operations against five initially planed). The geometrical specification (figure 1) is now respected.

If we want to improve this solution, it is necessary to take into account more driving parameters with more than four lobes defects. If we look at the modal decomposition of the optimal part, it can be observed that the most important defect corresponds to the mode L5. As it has been explained by the exploitation of the polynomial relation, there is no DCT coefficient that has an effect on it. A higher level of DCT (A3 and more) would decrease the residue.

## 3. Conclusion and outlook

In this paper, it is proposed a new approach for reducing the forming operations in the case of deep drawing. Thanks to a previous result, we have obtained a part with a minimal thickness reduction. This phenomenon corresponds to the main problem in the manufacture of this kind of part. From the initial progression, composed of five steps (blank cutting, forming, flanging and cutting), it is proposed to reduced it to only two (blank cutting and forming). For that, it was developed a compensation of the ear formation, in the aim to obtain at the end of the first forming operation a final product. A modal decomposition will allow us to characterize the ear shape. We defined particular polynomial models linking different blanks (size and shape) generated by discrete cosine transformation to the modal decomposition coefficients. Using these models, an optimization is possible in the aim to find the best blank which leads to minimize the ear apparition (by reducing by ten the ear amplitudes).

This approach allows to simplify the industrial forming progression drastically (from five to two) and to minimize the needed material.